\documentclass[journal]{IEEEtran}
\usepackage{multirow}
\usepackage{algorithm}
\usepackage[noend]{algpseudocode}
\usepackage{graphicx}
\graphicspath{ {images/} }
\usepackage{xcolor}
\usepackage{lettrine}
\usepackage{hyperref} 

\usepackage{amsmath}
\usepackage{amssymb}
\usepackage{booktabs}
\DeclareMathOperator*{\argmin}{min}

\newcommand\tabitem{\makebox[1em][r]{\textbullet~}}
\usepackage{pifont}
\newcommand{\cmark}{\ding{51}}
\newcommand{\xmark}{\ding{55}}

\definecolor{section}{RGB}{0, 0, 255}
\definecolor{subsection}{RGB}{138, 43, 226}

\ifCLASSINFOpdf
\else
\fi

\begin{document}

\title{Deep Perceptual Enhancement for Medical Image Analysis}
%
%
%

\author{S M A Sharif,~\IEEEmembership{}
        Rizwan Ali Naqvi,~\IEEEmembership{}
        Mithun Biswas,~\IEEEmembership{}
        and~Woong-Kee Loh~\IEEEmembership{}
\thanks{S M A Sharif and Mithun Biswas are with rigel-IT, Dhaka-1219 Bangladesh e-mail: (sma.sharif.cse@ulab.edu.bd; mithun.bishwash.cse@ulab.edu.bd).}
\thanks{Rizwan Ali Naqvi is with Department of Unmanned Vehicle Engineering, Sejong University, Republic of Korea e-mail: (rizwanali@sejong.ac.kr) }
\thanks{Woong-Kee Loh is with School of Computing, Gachon University, Seongnam, Republic of Korea e-mail: (wkloh2@gachon.ac.kr).}
\thanks{Manuscript received April 19, 2005; revised August 26, 2015. correspondence: (Rizwan Ali Naqvi; Woong-Kee Loh), equal contribution: (S M A Sharif; Rizwan Ali Naqvi) }}

%



\maketitle

\begin{abstract}
Due to numerous hardware shortcomings, medical image acquisition devices are susceptible to producing low-quality (i.e., low contrast, inappropriate brightness, noisy, etc.) images. Regrettably, perceptually degraded images directly impact the diagnosis process and make the decision-making manoeuvre of medical practitioners notably complicated. This study proposes to enhance such low-quality images by incorporating end-to-end learning strategies for accelerating medical image analysis tasks. To the best concern, this is the first work in medical imaging which comprehensively tackles perceptual enhancement, including contrast correction, luminance correction, denoising, etc., with a fully convolutional deep network. The proposed network leverages residual blocks and a residual gating mechanism for diminishing visual artefacts and is guided by a multi-term objective function to perceive the perceptually plausible enhanced images. The practicability of the deep medical image enhancement method has been extensively investigated with sophisticated experiments. The experimental outcomes illustrate that the proposed method could outperform the existing enhancement methods for different medical image modalities by 5.00 to 7.00 dB in peak signal-to-noise ratio (PSNR) metrics and 4.00 to 6.00 in DeltaE metrics. Additionally, the proposed method can drastically improve the medical image analysis tasks' performance and reveal the potentiality of such an enhancement method in real-world applications. Code Available: \url{https://github.com/sharif-apu/DPE_JBHI}.

\end{abstract}


\IEEEpeerreviewmaketitle

\section{ Introduction}

\IEEEPARstart{A}{cquisition} of medical images is known to be a strenuous process. The stochastic lighting condition and the subject motion (i.e., respiratory motion and physical movement) make the acquisition process substantially complicated \cite{sharif2020learning,ma2020cycle}. Arguably, a higher scanning time can address the problem that arises due to quantum inefficiency in stochastic lighting conditions to a small extend. Nonetheless, capturing images with such a configuration does not fit well in most common real-world scenarios. A higher scanning time increases the probability of producing visual artefacts like motion blurs and intensity inhomogeneity during the image acquisition \cite{yang2010medical,agarwal2018medical}. Hence, most medical imaging systems leverage the shorter scanning time to avoid such inevitable artefacts. Regrettably, the trade-off between shorter scanning time and artefacts drives the medical imaging system to another challenging dilemma and captures visually implausible low-quality images with sensor noise, insufficient luminance information, low contrast ratio, etc. \cite{sharif2020learning, agarwal2018medical}.

Contrarily, the visual quality (also referred to as {\it perceptual quality} in the rest of the paper) of images plays a crucial role in medical image analysis \cite{panayides2020ai,agarwal2018medical, yang2010medical,subramani2018mri,patel2013comparative}. The perceptual quality of an image considerably influences the decision taking manoeuvres of a medical expert and can directly impact the diagnosis process \cite{sagheer2020review,kollem2019review}. Such perceptual quality also determines the performance of computer-aided diagnosis (CAD) applications like segmentation \cite{pham2000current}, detection \cite{de2016machine}, recognition \cite{jebelli2018continuously}, etc. Despite the importance of perceptually plausible medical images, enhancing a degraded medical image always remains a challenging task. Even a minor misinterpretation of details while enhancing the sensitive medical image can drive towards catastrophe.  

A few works have addressed the challenge of medical image enhancement in recent times. These methods can be categorised into two major categories: traditional (non-learning-based) methods and learning-based methods. Typically, non-learning enhancement methods aim to improve the image quality by exploiting spatial and frequency transformation techniques, including histogram equalisation \cite{patel2013comparative}, histogram transformation \cite{jiang2015color}, edge extraction \cite{arivazhagan2007multi}, smooth filtering \cite{yang2014fast}, etc. Inversely, the learning-based methods learn to enhance the degraded medical image from a large set of image samples. Either way, the existing works on medical image enhancement aim at specific image manipulation tasks like contrast-enhancing \cite{huang2020differential,chondro2016effective}, intensity correction \cite{ma2020cycle,jiang2015color,patel2013comparative,yang2010medical}, and denoising \cite{sharif2020learning,gondara2016medical,jifara2019medical} separately rather than comprehensively improving the perceptual quality of low-quality images.  Subsequently, a comprehensive pipeline for perceiving plausible medical images is yet to explore.

Apart from medical imaging, the past decade has experienced a revolutionary takeover in different image manipulation tasks by deep learning. Several recent learning-based methods illustrate tremendous results in low-quality image enhancement, including the images captured with stochastic lighting conditions \cite{lore2017llnet,wei2018deep, guo2020zero}. Several deep methods demonstrate state-of-the-art performance in reconstructing high dynamic range (HDR) images from low dynamic range (LDR) inputs \cite{hdr2021ntire,sharif2021two}. Even a few methods have illustrated the capability of mimicking complicated image signal processing (ISP) pipeline \cite{sharif2021BJDD, ignatov2020replacing} by leveraging deep learning. Although medical images are substantially different from their non-medical counterparts, the success of these non-medical image manipulations inspired this study to tackle the challenge of medical images enhancement comprehensively. Please see Table. \ref{comIntro} for a brief comparison between proposed method and existing works.

\begin{table*}[!htb]
\caption{Brief comparison between existing and proposed perceptual medical image enhancement method}
\centering
\scalebox{.85}{\begin{tabular}{lp{3cm}p{3cm}p{4.8cm}p{4.8cm}}
\toprule[0.4pt]\toprule[0.4pt]
\textbf{Category}                                & \textbf{Approach}         & \textbf{Method}                                                       & \textbf{Strength}                                                                                                                                                                           & \textbf{Weakness}                                                                                                         \\ \toprule
\multirow{2}{*}{\textbf{Medical Image}} & Non learning              & Spatial and frequency transformation                                   & \begin{tabular}[c]{@{}l@{}} \tabitem Do not require large dataset \\ \tabitem Can perform specific enhancement tasks  \end{tabular}                                                   & \begin{tabular}[c]{@{}l@{}} \tabitem Can not handle diverse data sample \\ \tabitem  Unable to perform comprehensive \\ enhancement\end{tabular} \\
                                        & Learning-based            & Learns from medical images                                            & \begin{tabular}[c]{@{}l@{}} \tabitem Can outperform non-learning methods \\ \tabitem Promising performance in specific \\ enhancement tasks\end{tabular}                                            & \begin{tabular}[c]{@{}l@{}} \tabitem Need a large number of data sample \\ \tabitem Optimised for specific enhancement task\end{tabular}     \\ \toprule
\textbf{Computer Vision}                & Image Enhancement         &  Utilises optimisation or deep learning for enhancing non medical images & \begin{tabular}[c]{@{}l@{}} \tabitem Can perform comprehensive enhancement \\ \tabitem Can handle diverse data sample\end{tabular}                                                                             & \begin{tabular}[c]{@{}l@{}} \tabitem Need large number of data samples \\ \tabitem Not been studied for medical images\end{tabular}          \\ \toprule
\textbf{Proposed}                       & Medical Image Enhancement & Leverage perceptual optimisation                                      & \begin{tabular}[c]{@{}l@{}} \tabitem Designed for medical images enhancement \\ \tabitem Accelerate Computer aided diagnosis (CAD) applications and \\diagnosis process \\ \tabitem Lightweight and can handle real-world \\medical images\end{tabular} & \tabitem Learns from sythesized data samples   \\ \toprule[0.4pt]\toprule[0.4pt]                                                                                      
\end{tabular}}
\label{comIntro}
\end{table*}

This study proposes a novel learning-based method to tackle the challenge of enhancing low-quality medical images.Thus, the enhanced medical images can accelerate the performance of CAD applications and help out medical image experts in medical image analysis tasks. To the best concern, this is the first work in the open literature that learns perceptual enhancement of medical images, including intensity normalisation, contrast correction, noise suppression, etc., comprehensively by incorporating a convex set of medical images. The proposed network appropriates a residual feature gatting \cite{yu2019free} and a residual block \cite{he2016deep} in a deep encoder-decoder-like \cite{cho2014learning,gao2019deep} structure. Here, the residual blocks allow the deep network to perceive a deeper architecture and feature gates imply propagating important features to reduce visual artefacts from the enhanced images. Additionally, the proposed method also presents a multi-term perceptual objective function for perceiving visually admissible images. The experimental results illustrate the proposed deep method can handle a diverse range of medical image modalities and considerably outperform the state-of-the-art enhancement methods in image evaluation metrics. Depending on the image modalities, the proposed method could improve 5.00 to 7.00 dB in PSNR metrics and 4.00 to 6.00 in DeltaE metrics compared to its counterparts. The feasibility of the proposed method has also extensively studied by incorporating it into different medical image analysis tasks.

\begin{figure*}[!htb]
\centering
\includegraphics[width=16cm,keepaspectratio]{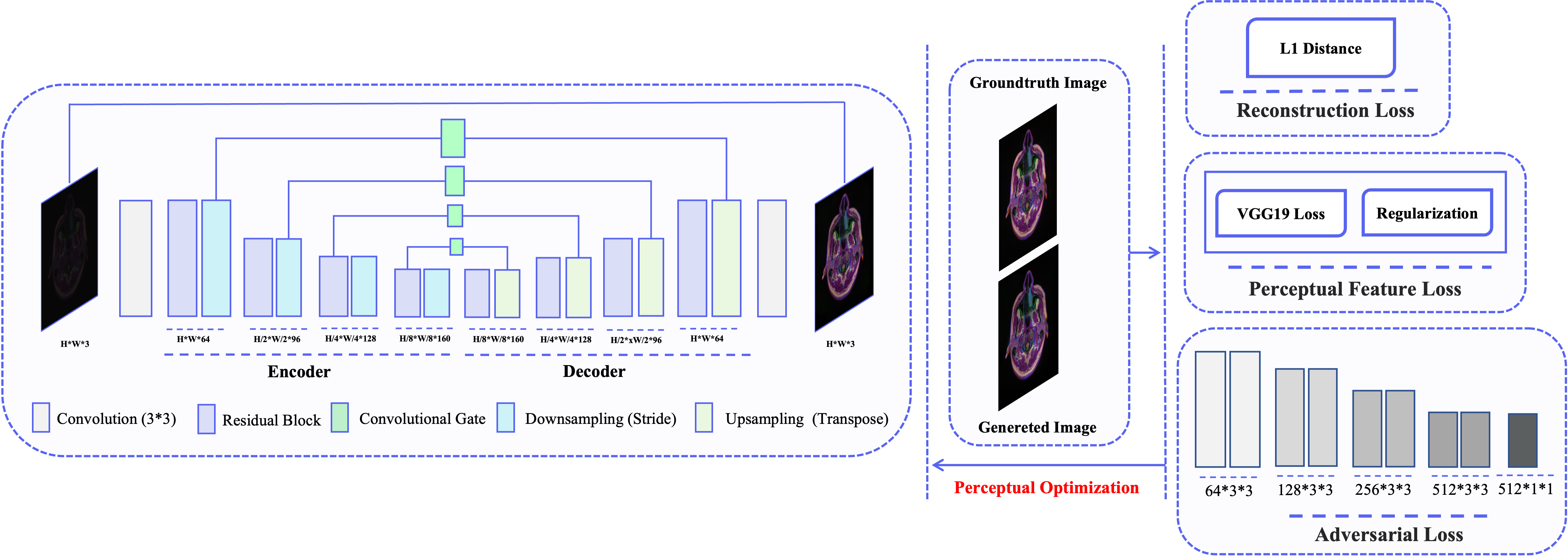}
\caption{ The overview of the proposed deep perceptual enhancement method. The proposed network incorporates residual blocks and features a gating mechanism in an encoder-decoder like structure to reduce visual artefacts. Also, the proposed deep network has guided by the multi-term objective function, which aims to enhance the perceptual quality of low-quality images. }
\label{network}
\end{figure*}

\section{Methods}
This study introduces a learning-based deep method to enhance low-quality medical images captured with multiple medical image acquisition devices. This section explains the network architecture, optimisation scheme, data simulation process and implementation strategies in detail.

\subsection{Deep Perceptual Network}
The proposed method considers perceptual enhancement of medical images as an image-to-image translation task. Hence, it maps a low-quality image ($I_U$) as $\mathrm{G}: I_U \to  I_R$. Where mapping function ($\mathrm{G}$) learns to generate perceptually enhanced image ($I_R$) as $I_R \in [0,1]^{H \times W \times 3}$. $H$ and $W$ represents the height and width of the input as well as output images. Fig. \ref{network} shows the overview of the proposed deep method.

As shown in  Fig. \ref{network}, the proposed deep network has designed as a fully convolutional encoder-decoder deep network \cite{gao2019deep,cho2014learning} with gated skip connections. The first layer of the generator takes an input image ($I_U$) and turns it into a 64 depth feature map. The input convolutional layer comprises a kernel size of $3 \times 3$, padding of 1, and stride of 1. An encoder follows the input layer with four consecutive feature levels, where feature depths are alternated as $d ={64, 96, 128, 196 }$. Each of these feature levels includes a residual block, which maps an input feature map $X$ as:

\begin{equation}
X^{\prime} = \mathrm{M}(X) + X
\end{equation}

Here, $\mathrm{M}(\cdot)$ comprises two convolution layers, where each convolution comprises a kernel size of $3 \times 3$, padding of 1, and stride of 1, and the first layer is activated with a PRelU activation. The output feature dimension of a residual block remains unchanged as its input (i.e., 64 in the first feature level).  Each residual block in the encoder is followed by a convolutional downsampling layer as follows:

\begin{equation}
F_{\downarrow} = C_{\downarrow} (X_{0})
\end{equation}

Here, $C_{\downarrow}$ represents $3 \times 3$ convolution operation with stride=2.

A convolutional decoder follows the encoder portion of the network. The decoder leverages the same number of feature levels as the encoder and comprises an upsampling layer right after every residual block, as shown in Fig. \ref{network}. Here the upsampling has been perceived  as follows:

\begin{equation}
F_{\uparrow} = C_{\uparrow} (X_{0})
\end{equation}
 Here, $F_{\uparrow}$ represents transpose convolution operation \cite{dumoulin2016guide}.

The decoder portion is followed by the final convolutional layer, which outputs the three-channel enhanced image by incorporating a convolutional kernel size of $3 \times 3$, padding of 1, the stride of 1. The final output layer has activated with a tanh activation to output the final images in a range of $[0,1]$.

Apart from leveraging encoder-decoder architecture, the proposed method also utilised a contextual gating mechanism  \cite{yu2019free} instead of traditional residual skip connections \cite{he2016deep}. Here, the contextual gate aims to propagate the important feature by pruning the trivial features as follows:
 
\begin{equation}
A_{m,n} = \sum_{m=1}^{H}\sum_{n=1}^{W} W_{a}.I 
\end{equation}

\begin{equation}
B_{m,n} = \sum_{m=1}^{H}\sum_{n=1}^{W} W_{b}.I 
\end{equation}

\begin{equation}
O_{m,n} = \phi(A_{m,n})	\odot \delta(B_{m,n})
\end{equation}
 
Here, $ \phi $ and $\delta$ present the LeakyReLU and sigmoid activations. $W_{a}$, $W_{b}$ represent convolutional operations and $\odot$ presents multiplication.

\subsection{Perceptual Optimization}

The proposed deep model ($\mathrm{G}$) has optimised with a novel multi-term perceptual loss function. For a given  training set $\{ I_U^t, I_G^t \}_{t=1}^P$ consisting of $N$ image pairs, the training process aims to minimise the objective function described as follows:

\begin{equation}
 W^\ast = \arg{\argmin_W}\frac{1}{P}\sum_{t=1}^{P}\mathcal{L}_{\mathit{P}}(\mathrm{G}(I_U^t), I_G^t)
 \label{fLoss}
\end{equation}
 
Here, $\mathcal{L}_{\mathit{P}}$ denotes the perceptual loss, and $W$ presents the parameterised weights of the proposed deep model. The proposed perceptual loss improves the perceptual quality of low-quality medical images. 

\textbf{Reconstruction loss.} Typically, learning-based image-to-image translation methods utilise a pixel-wise L1 or an L2 distance to obtain a refine the output. However, several recent studies  \cite{schwartz2018deepisp, sharif2021BJDD, sharif2021sagan} have reported that L2-loss are prone to produce smoother images due to their direct relationship with the PSNR \cite{huynh2008scope}. Smoother images are considerably similar to homogenous blurs and can make the diagnosis process strenuous for experts and CAD applications. To avoid such unwanted drawbacks, this study exploited an L1 norm as a reconstruction loss during the training phase:

\begin{equation}
 \mathcal{L}_{\mathit{R}} = \parallel I_G-I_R \parallel_1
\end{equation}

Here, $I_R$ presents the enhanced output of $\mathrm{G}$ and $I_G$ presents the reference image.

\textbf{Regularised Feature loss.} VGG-19 network-based feature loss encouraged a generated image to have a similar perceptual feature representation as its reference images \cite{johnson2016perceptual, wang2018esrgan}. Subsequently, such feature loss can guide a deep model to perceive visually plausible details. However, an empirical setup of a pre-trained VGG-19 network with the ReLU activation can drive a deep network to produce inconsistent brightness and contrast \cite{sharif2021BJDD,wang2018esrgan}.   A recent study has introduced a variant of feature loss known as regularised feature loss \cite{sharif2021BJDD} to counter such contradictory consequences. This study adopted the regularised feature loss instead of typical feature loss to enhance the details of low-quality medical images as follows:

\begin{equation}
 \mathcal{L}_{\mathit{RFL}} = \lambda_R \times
 \frac{1}{H_j W_j  C_j}\parallel \psi(I_G) - \psi(I_R) \parallel_1
\end{equation}

Here, $\lambda_R$ represents total variant regularise and $\psi$ denotes the pre-trained VGG network.

\textbf{Adversarial loss.} Adversarial guidance has proven to be applicable for producing colourful, natural-looking images with dense structure information \cite{ignatov2017dslr}.  This study incorporates a  variant of the generative adversarial network (GAN) \cite{goodfellow2014generative} known as conditional GAN (cGAN) \cite{mirza2014conditional} for perceiving colour consistency and dense texture representation in enhanced medical images. Here, the cGAN adversarial loss aims to minimise the cross-entropy loss of its generator as follows: 
 
\begin{equation}
 \mathcal{L}_{\mathit{G}}= - \sum_{t} \log D(I_R, I_G)
\end{equation}
 
  Here, $\mathrm{D}(\cdot)$ presents the discriminator and cGAN based adversarial guidance intends to maximise it as $\mathbb{E}_{X, Y} \big[\log D\big(X, Y\big) \big]$. This study develops a discriminator with nine convolutional layers. The first eight layers of the conditional discriminator comprise a kernel size of $3 \times 3$ followed by a swish activation. And the final convolution operation incorporates a kernel size of $1 \times 1$ and is activated by a sigmoid function. Notably, every $(2n-1)^{th}$ layer of the discriminator alters the depth and spatial dimension by a factor of 2, as shown in fig. \ref{network}. 

\textbf{Perceptual Optimisation.} The perceptual optimisation $\mathcal{L}_{\mathit{P}}$ has obtained by combining multi-term losses as follows:
 
\begin{equation}
 \mathcal{L}_{\mathit{P}}= \mathcal{L}_{\mathit{R}} + \mathcal{L}_{\mathit{RFL}} +  \lambda_{G}.\mathcal{L}_{\mathit{G}}
\end{equation}
 
Here, $\lambda_{G}$ presents the loss regulators, and it has been set as $\lambda_{G}$ = 1e-4 for stabilizing adversarial loss.

\begin{figure}[!htb]
\centering
\includegraphics[width=8.5cm,keepaspectratio]{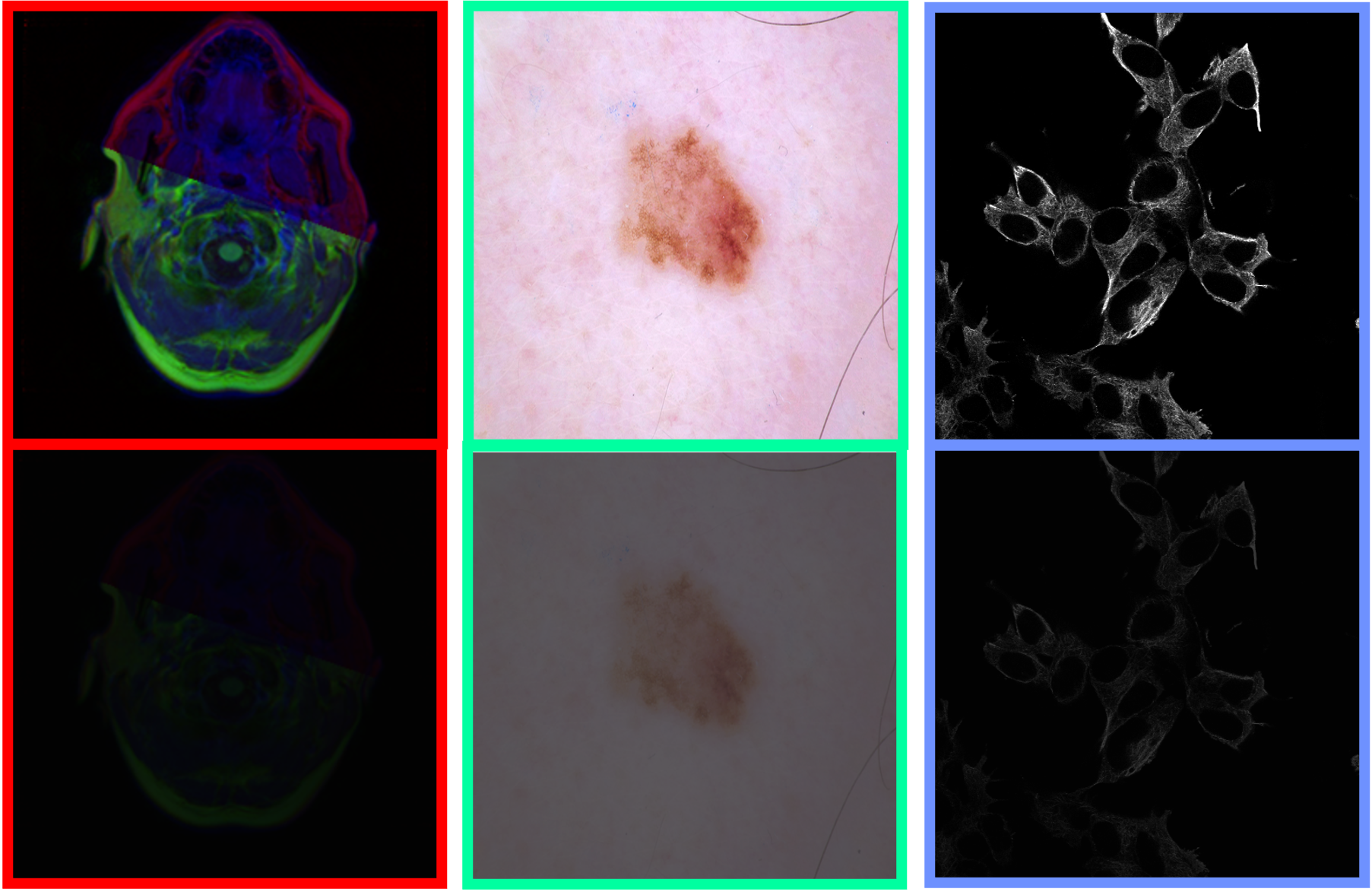}
\caption{ Pair-wise image generation obtained by the proposed method. In every pair, top: reference image, and bottom: simulated low-quality image. }
\label{dataset}
\end{figure}

\subsection{Dataset Preparation}

Collecting a substantial amount of medical images is always considered a challenging task \cite{arsalan2017deep, sharif2020learning}. Most importantly, there is no specialised dataset available in the open literature for studying the perceptual enhancement of medical images. Hence, this study collected over 30,000  medical images from the open web and introduced a novel algorithm for synthesising the low-quality medical images for pair-wise training. The collected images include diverse data samples captured with multiple medical image acquisition devices. The collected images divided into three major categories: i) Radiology (i.e., magnetic resonance images (MRI) \cite{buda2019association}, X-ray \cite{irvin2019chexpert}), ii) Dermatology (i.e., skin images \cite{rezvantalab2018dermatologist}), iii) Microscopy (i.e., protein atlas \cite{uhlen2010towards}). It is worth noting that the proposed method can handle both monochrome and RGB images. Similar to previous studies \cite{sharif2020learning,khanh2020enhancing}, this study leveraged the MRI images with pseudo-colour for qualitative comparison. It allowed us to illustrate the meaningful improvement in terms of colour consistency across different images. 

Apart from that, collected data samples comprised a significant amount of low-quality images and required enhancement. It has been found that downgrading such low-quality images makes the model training ambiguous. This study proposed a low-quality medical image simulation algorithm to address such deficiencies. As the Algorithm. \ref{dataSyth}  shows that the proposed method synthesised the low-quality medical images by altering brightness, contrast and adding random noises \cite{sharif2020learning}. Here, the algorithm estimates the luminance information by converting RGB images into Lab colour space. The luminance threshold is intended to eliminate predetermined low-quality medical images by observing their intensity while synthesising the pair-image dataset. Later, the target image samples were contaminated by random noise distribution with a standard deviation of $ \sigma = 10$. Additionally, the exposure-related parameters like brightness, contrast, etc., were randomly modified between $[0, 0.8]$. Fig. \ref{dataset} illustrates the  data samples from the synthesised dataset. The synthesised dataset incorporates 15181 image samples for training and 3,000 images for performance evaluation (i.e., validation and testing).

\begin{algorithm}
\caption{Low-quality medical image simulation}\label{alg:euclid}
\begin{algorithmic}[24]
\State $B_T$ = 100 - threshold of brightness
\State $R_{max}$ = 0.8 - maximum value of random range 
\State $\sigma$ = 10 - maximum standard deviation of Gaussian noise
\Procedure{SIMULATION}{$I_R,B_T, R_{max}, \sigma$}
\State $I_T\gets Lab(I_R)$ 
\State $I_L\gets I_T[:, : , 0:1]$ 
\State $k\gets 0$
\For{$i \gets 1$ to $N$}                    
    \For{$j \gets 1$ to $M$}              
        \If{$I_L[i,j]$ $>$ $0$ }
            \State {$L_T[k]$ $\gets$ $I_L[i,j]$}
            \State $k\gets k+1$
        \EndIf
    \EndFor
\EndFor

\State $L\gets average(L_T)$
\If{$L$ $>$ $B_T$}

    \State $R_N\gets uniform(0, \sigma)$
    \State $N\gets noise(I_R, R_N)$ 
    
    \State $I_N\gets I_R \times N$

    \State $R_C\gets uniform(0, R_{max})$
    \State $I_C\gets contrast(I_N, R_C)$ 
    
    \State $R_B\gets uniform(0, R_{max})$
    \State $I_B\gets brightness(I_N, R_B)$ 

    \State $I_D\gets I_N \times I_C \times I_B$
\EndIf

\State \textbf{return} $I_D$
\EndProcedure
\end{algorithmic}
\label{dataSyth}
\end{algorithm}

\subsection{Training Details}

\begin{figure*}[!htb]
\centering
\includegraphics[width=14cm,keepaspectratio]{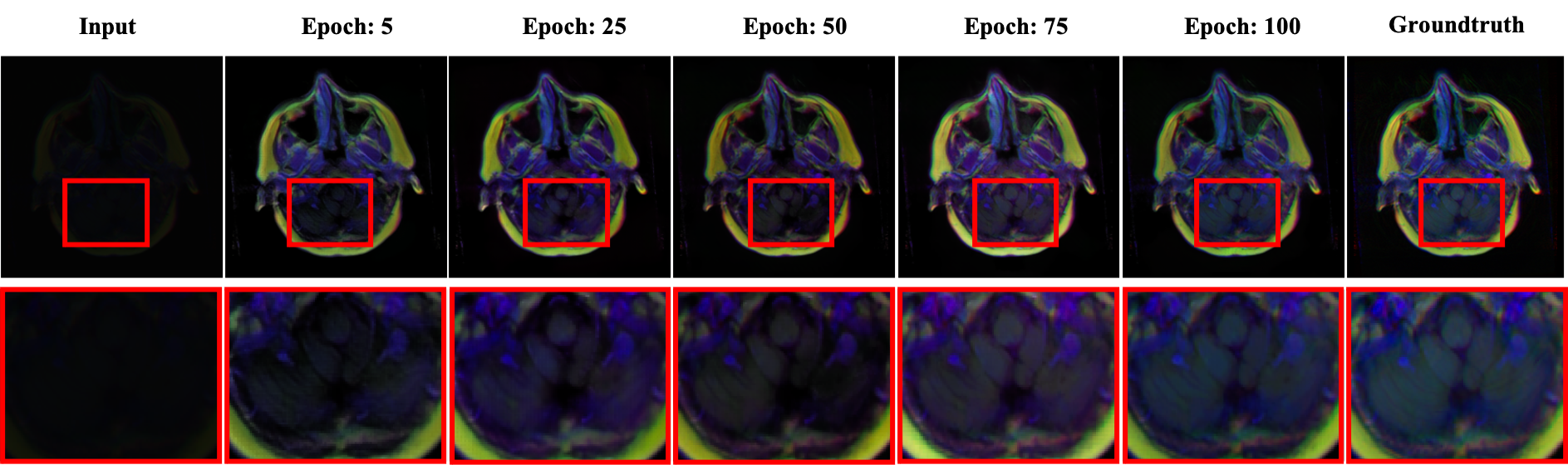}
\caption{ Visualization of the training phase. The proposed method has trained until the model converges with the given images. }
\label{training}
\end{figure*}

The proposed deep perceptual enhancement method implemented with the PyTorch \cite{pytorch}.  The proposed model is fully convolutional and it can infer with any dimensioned images during the training and testing phases. However, for making the training process generic, training images were resized into $128 \times 128 \times 3$ . Later, the model was tested images with their actual dimensions. The network was optimised by an  Adam optimiser \cite{kingma2014adam} with the hyperparameters  $\beta_1 = 0.9$ and $\beta_2 = 0.99$. Additionally, the model trained for 100 epochs  with a constant batch size of 24.  Fig. \ref{training} shows the result obtained during the training phase. It took around 24 hours to converge the proposed model with a constant learning rate of 5e-4. All experiments were conducted on a hardware setup, including an AMD Ryzen 3200G central processing unit (CPU) clocked at 3.6 GHz, a random-access memory of 16 GB, and An Nvidia Geforce GTX 1060 (6GB) graphical processing unit (GPU).

\section{Results}
The feasibility and applicability of a deep learning-based medical image enhancement method have extensively studied throughout this study. This section details the results of the proposed medical image enhancement method, its feasibility in the medical image analysis task and its practicability in a real-world application.

\subsection{Evaluation and Comparison}
The performance of the enhancement methods has evaluated with medical images acquired by multiple medical modalities. The evaluation has conducted both subjectively and objectively to analyse the performance of the proposed deep method in such diverse data space. Simultaneously, the performance of the proposed method has compared with the state-of-the-art enhancement methods.  It is worth noting a comprehensive medical image enhancement has never been done before. Therefore, the non-medical image enhancement methods have been considered for comparison. The comparing enhancement methods are as follows:  i) LIME \cite{guo2016lime}, ii) DUAL \cite{zhang2019dual}, and iii) RetnixNet \cite{wei2018deep}. Apart from that, a recent medical image enhancement method (DRAN \cite{sharif2020learning}) has been studied to illustrate the performance of such a deep method for comprehensive medical image enhancement.  DRAN \cite{sharif2020learning} tackles the challenges of medical image denoising. For a fair evaluation, all enhancement methods were studied under the same data samples. Notably, such dense comparison allows this study to analyse the performance of different image enhancement methods in medical image enhancement.

\subsubsection{Quantitative Evaluation}

\begin{table*}[!htb]
\caption{Quantitative evaluation of proposed method and existing image enhancement methods. The proposed method outperforms its counterparts for medical image enhancement with a substantial margin. }
\centering
\begin{tabular}{lllllllll}
\toprule[0.4pt]\toprule[0.4pt]
\multirow{2}{*}{\textbf{Method}}         & \multicolumn{2}{l}{\textbf{Radiology Images}} & \multicolumn{2}{l}{\textbf{Dermatology Image}} & \multicolumn{2}{l}{\textbf{Microscopic Images}} & \multicolumn{2}{l}{\textbf{Average}} \\ \cmidrule{2-9} 
                                & \textbf{PSNR $\uparrow$}      & \textbf{DeltaE}  $\downarrow$   & \textbf{PSNR $\uparrow$}          & \textbf{DeltaE $\downarrow$ }        & \textbf{PSNR $\uparrow$}          & \textbf{DeltaE $\downarrow$ }         & \textbf{PSNR $\uparrow$ }     & \textbf{DeltaE $\downarrow$ }   \\ \toprule
LIME \cite{guo2016lime}                            & 19.14              & 9.34                & 19.32                  & 10.32                  & 17.07                  & 7.83                    & 18.51             & 9.16              \\ 
DUAL \cite{zhang2019dual}                            & 21.69              & 6.94                & 16.70                  & 13.26                  & 25.21                  & 3.05                    & 21.20              & 7.75              \\ 
RetnixNet \cite{wei2018deep}                      & 20.71              & 10.89               & 18.28                  & 11.32                  & 22.10                  & 8.91                    & 20.36             & 10.37             \\ 
DRAN \cite{sharif2020learning}                      & 27.47              & 4.23               & 18.77                  & 9.14                  & 30.38                  & 1.31                    & 25.24             & 4.89             \\ 
\textbf{Proposed} & \textbf{29.04}     & \textbf{3.21}       & \textbf{23.11}         & \textbf{6.36}          & \textbf{30.69}         & \textbf{1.11}           & \textbf{27.61}    & \textbf{3.56}     \\ \toprule[0.4pt]\toprule[0.4pt]
\end{tabular}
\label{quanRes}
\end{table*}

Quantitative evaluation plays a significant role in determining the performance gain of any image enhancement method. Hence, the performance of the proposed method has quantitatively evaluated and compared with its counterparts.  The enhancement performance has summarised with two well-known evaluation metrics: PSNR \cite{huynh2008scope} and DeltaE \cite{luo2001development}. Both evaluation metrics evaluate the enhanced image by comparing it with the corresponding ground-truth image, like the human visual system. The PSNR metrics intend to illustrate the performance gain in terms of noise suppression, while the DeltaE metric calculates the perceptual enhancement obtained by the respective enhancement methods. The higher value of PSNR illustrates better enhancement, and the lower value of DeltaE indicates better perceptual quality.  Table. \ref{quanRes} demonstrates the enhancement performance of all comparing methods for different medical image categories. Here, the residual block and gating mechanism allows the proposed method to reduce visual artefacts, while perceptual optimisation encourages the proposed model to produce visually plausible images. As a consequence, it substantially outperforms the existing image enhancement methods in all evaluating criteria.  On average, the proposed method can gain around 5.00 to 7.00 dB in PSNR metrics and 4.00 to 6.00 in DeltaE metrics to exceed its counterparts.

\subsubsection{Qualitative Evaluation}

\begin{figure*}[!htb]
\centering
\includegraphics[width=15cm,keepaspectratio]{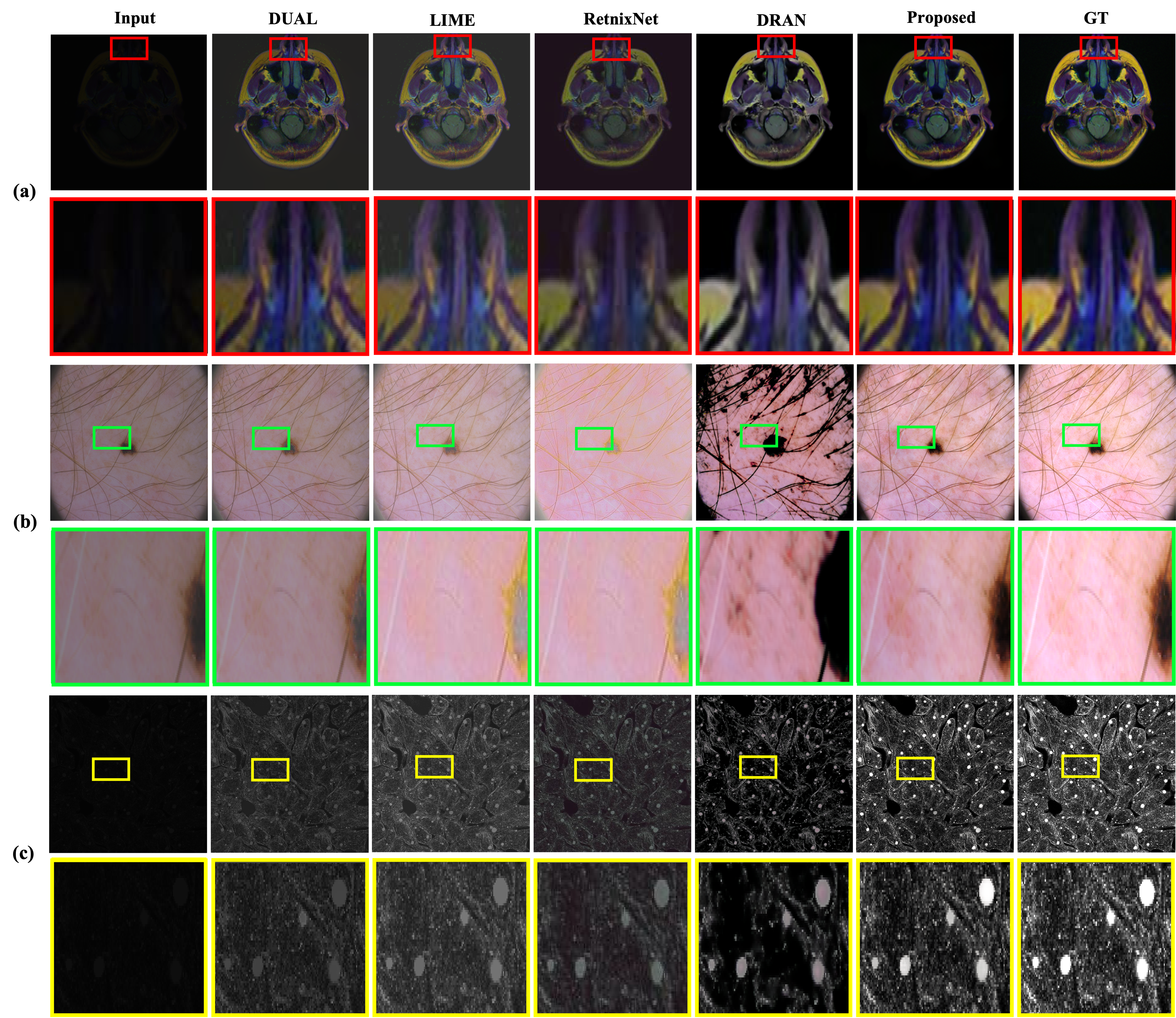}
\caption{Qualitative evaluation of proposed method and existing image enhancement methods. The proposed method illustrates consistency over every comparing medical image category. In every category, left to right: input (low-quality) images, results of LIME \cite{guo2016lime}, DUAL \cite{zhang2019dual}, RetnixNet \cite{wei2018deep}, DRAN \cite{sharif2020learning},  Proposed method and, ground truth images. (a) Radiology Images. (b) Dermatology Image. (c) Microscopic Images.}
\label{qualComp}
\end{figure*}

Qualitative evaluation intends to illustrate a visual representation of perceptual improvement obtained by the enhancement methods. Fig. \ref{qualComp} depicts the perceptual enhancement perceived by the proposed method and comparing state-of-the-art image enhancement methods. It is visible that the proposed method generates the most natural-looking (i.e., not over-processed) enhanced images among all comparing enhancement methods. Also, the enhancement performance of the proposed method is consistent for all medical image modalities. Contrarily, the existing image enhancement methods tend to over-processed the input images. Which instantaneously pushes them to produce visually disturbing artefacts with inconsistent luminance information and contrast. 

\subsubsection{Inference Time}

\begin{table}[!htb]
\caption{Inference time (in seconds) required for enhancing low-quality medical images with the proposed method.}
\centering
\begin{tabular}{lllllll}
\toprule[0.4pt]\toprule[0.4pt]
\multirow{3}{*}{\textbf{Model}} & \multicolumn{6}{l}{\textbf{Resolution}}                                                                                                \\ \cmidrule{2-7} 
                                & \multicolumn{2}{l}{\textbf{256 * 256 * 3}} & \multicolumn{2}{l}{\textbf{512 * 512 * 3}} & \multicolumn{2}{l}{\textbf{1024 * 1024 * 3}} \\ \cmidrule{2-7} 
                                & \textbf{CPU}         & \textbf{GPU}        & \textbf{CPU}         & \textbf{GPU}        & \textbf{CPU}          & \textbf{GPU}         \\ \toprule
LIME                            & 0.75                 & -                   & 4.1                  & -                   & 26.42                 & -                    \\
DUAL                            & 1.6                  & -                   & 7.9                  & -                   & 52.99                 & -                    \\
RetnixNet                       & 0.93                 & 0.05                & 3.45                 & 0.15                & 18.49                 & -                    \\
DRAN                            & 0.86                 & 0.015               & 3.25                 & 0.03                & 13.9                  & 0.06                 \\
\textbf{Proposed}               & \textbf{0.45}        & \textbf{0.01}       & \textbf{1.65}        & \textbf{0.02}       & \textbf{6.85}         & \textbf{0.04}        \\
\toprule[0.4pt]\toprule[0.4pt]
\end{tabular}
\label{inference}
\end{table}

Inference time is considered one of the most crucial parts of a deep model and quantifies the ability of a specific model's processing time in real hardware. Despite illustrating a significant impact in different medical image analysis approaches, the proposed generator comprises only 4,486,987 trainable parameters. Such compact nature allows the proposed architecture to achieve a faster inference time compared to its counterparts. Table. \ref{inference} illustrates the performance comparison of different image enhancement methods and the proposed perceptual deep model. It is worth noting that even without a dedicated tensor computation engine (i.e., GPU), the proposed method can enhance large dimensioned medical images in a couple of seconds and outperform the existing methods. The inference time of the proposed method can be substantially decreased with a dedicated tensor computation engine. It is worth noting the proposed method has been designed as an end-to-end network. Therefore, the network does not require any additional pre or post-processing operation. Thus, the computational time is intended to be unchanged for a similar setup.

\subsection{Medical Image Analysis}
The motive of perceptual enhancement for medical image analysis is mainly two-fold. First, help the medical experts in the diagnosis process for patient treatment by offering visually plausible images. Second, allow the CAD systems to achieve an efficient performance. This study reveals the importance of perceptual enhancement by incorporating medical experts and CAD applications with sophisticated experiments.

\subsubsection{Expert Preference}

In most cases, medical experts such as medical practitioners and doctors are used to instantiate the diagnosis process by directly observing the medical images  \cite{sharif2020learning}. An expert study is conducted to examine the practicability of a perceptual medical image enhancement method. The expert study incorporated ten medical image experts aged between 30 to 50. 20 medical image sets, including 08 sets of MRI images, 08 sets of microscopic images, and 04 sets of dermatology images were presented to these experts. Each of these image sets comprised the input low-quality image and the enhanced results obtained by the image enhancement methods. The experts were invited to rank the given samples according to assessment criteria such as detail, colour, contrast, and brightness. Expert assessment results were applied to the Bradley-Terry model \cite{hunter2004mm,huang2006generalized} for summarising the expert preferences with a ranking system.

 Using the Bradley-terry model, the Proposed model along with the comparing models have been subjectively ranked. Here, the Bradley-Terry model performs as a probabilistic model to implement paired comparison. Later, the Newton-Raphson method \cite{akram2015newton} has been used to get the log-likelihood of the pairwise comparison observations. Note that, we have 6 × 6 matrices including all methods and Low-quality images (input). Six histograms have presented in Fig. 5, which shows the rank distribution of every comparing method over 20 medical image sets. Histograms are presented in a way that the best performing method is in a left position and gradually the later performing methods are positioned to the right. The proposed method has eight images in rank 1, five in rank 2, two in rank 3, and left the later ranks empty. It significantly outperformed the existing enhancement methods in the expert preference study. Also, it reveals that the medical image experts preferred to utilize the enhanced images over the low-quality variants for diagnosis purposes.

\begin{figure*}[!htb]
\centering
\includegraphics[width=15cm,keepaspectratio]{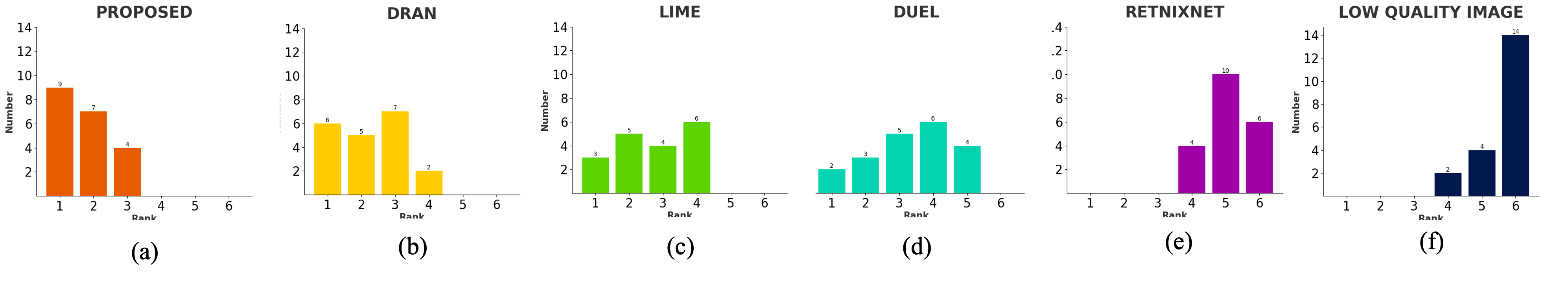}
\caption{ Summary of expert preference. The medical image experts preferred the proposed method over comparing methods. (a) Proposed method. (b) DRAN \cite{sharif2020learning} (c) LIME \cite{guo2016lime}. (d) DUAL \cite{zhang2019dual}. (e) RetnixNet \cite{wei2018deep}. (f) Low-quality medical image. }
\label{exPre}
\end{figure*}

\subsubsection{Computer Aided Diagnosis}
In recent times, CAD applications have gained popularity due to their widespread usage. Typically, a CAD system aims to detect trivial abnormalities by finding architectural distortion and prediction, which a human professional fails to discover with the naked eyes \cite{halalli2018computer,neofytou2014computer}. The perceptual quality of a medical image can considerably impact the performance of any CAD system.  Despite the in-network processing to component brightness and contrast insufficiency, CAD applications often suffer from low-quality medical images. Contrarily, the proposed method aims to accelerate the performance of such CAD applications by enhancing low-quality medical images with a small overhead cost.

\begin{figure}[!htb]
\centering
\includegraphics[width=9cm
,keepaspectratio]{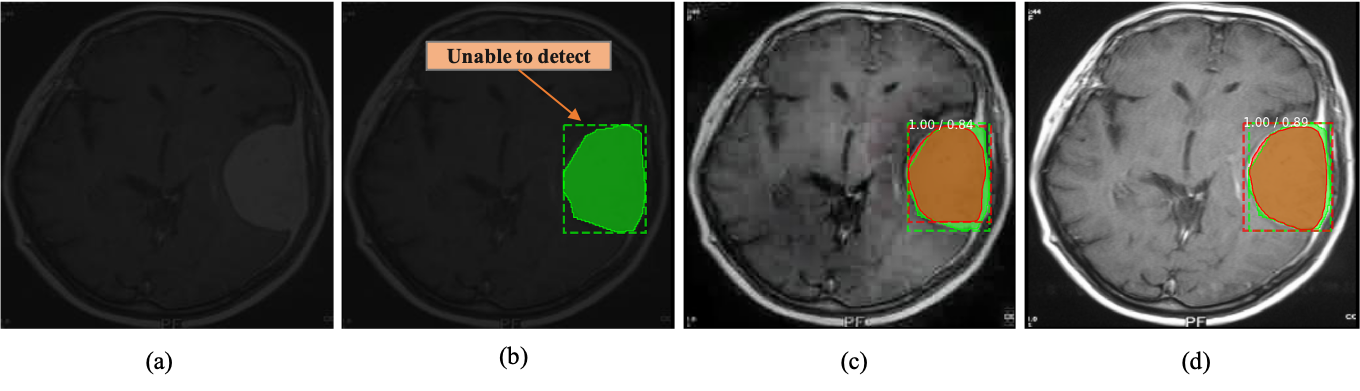}
\caption{ Abnormalities detection on brain MRI images. The low-quality medical images significantly deteriorate the performance of a detection method. The proposed method can improve the perceptual quality of a low-quality medical image and drastically improve the performance of the detection method. Here, green boxes indicate the ground-truth region, and deep-yellow boxes represent the detected area. (a) low-quality image. (b) low-quality + Mask R-CNN \cite{he2017mask}. (c) Enhanced image obtained by proposed method + Mask R-CNN \cite{he2017mask}. (d) Reference image  + Mask R-CNN \cite{he2017mask}.   }
\label{detection}
\end{figure}

\textbf{Abnormality Detection.} Abnormality detection analyses and predicts the abnormalities by observing the oversights of given images \cite{de2016machine}. Hence, a low-quality medical image can misinterpret a detection system by incorporating false positive or negative predictions, as shown in Fig. \ref{detection}. Here, the abnormalities (i.e., tumour) detection has performed with the state-of-the-art Mask R-CNN \cite{he2017mask} in MRI images. It can be visible that even a well-recognised detection method failed to identify the abnormalities in a low-quality image. Inversely, the proposed method can enhance the perceptual quality of the given low-quality medical image and dramatically accelerates the performance of the detection method.

\begin{figure}[!htb]
\centering
\includegraphics[width=9cm,keepaspectratio]{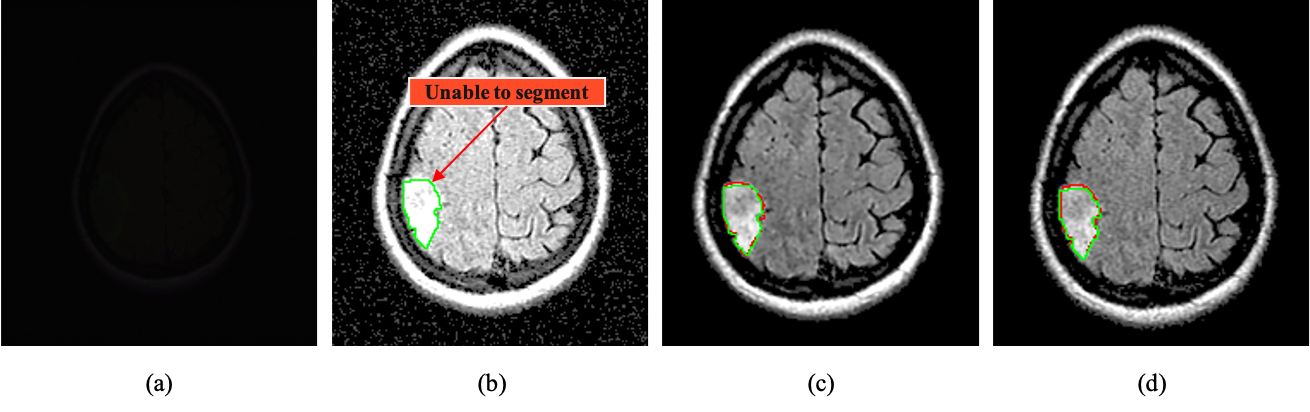}
\caption{ Abnormalities Segmentation on brain MRI images. The low-quality medical images impact the performance of any segmentation method. Notably, the primitive preprocessing also may not be sufficient to accelerate the performance of the segmentation methods. The proposed method can drastically accelerate the performance of the segmentation method by producing artefact-free plausible images. Here, green boxes indicate the ground-truth region, and red boxes represent the detected area. (a) low-quality image. (b) low-quality + U-Net \cite{ronneberger2015u}. (c) Enhanced image obtained by proposed method + U-Net \cite{ronneberger2015u}. (d) Reference image  + U-Net \cite{ronneberger2015u}. }
\label{segmentation}
\end{figure}

\textbf{Abnormalities Segmentation.} Abnormalities segmentation is known to be one of the most prominent CAD applications to date. Regrettably, such segmentation methods also suffer from low-quality images. Fig. \ref{segmentation} depicts the impact of the proposed method in improving the performance of such segmentation methods. Here, segmentation has performed on brain MRIs using well-known U-Net architecture \cite{ronneberger2015u}. It is worth noting that the original segmentation work itself performs processing to compensate perceptual degradation for accelerating its performance. However, their method miscalculates the perceptual attributes of the given image and substantially boosts the sensor noises. Such misinterpretation drives that segmentation method to unsatisfactory performance. Oppositely, the proposed method can enhance the low-quality MRI images by preserving natural attributes and dramatically increase the performance of the segmentation method.

\begin{figure*}[!htb]
\centering
\includegraphics[width=15cm,keepaspectratio]{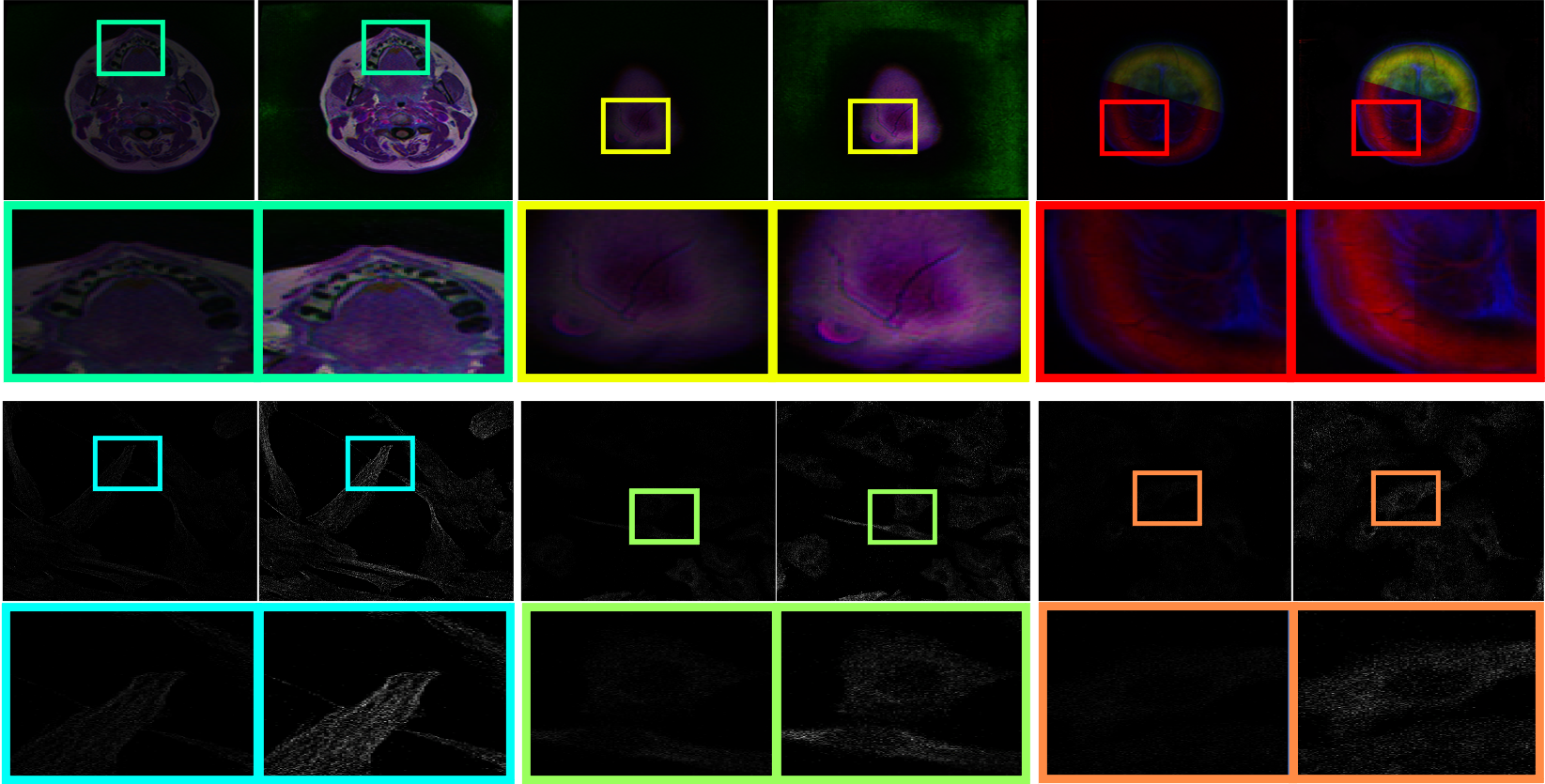}
\caption{ Real-world medical image enhancement with the proposed method. Despite being trained with synthesized images, the proposed method can handle the real-world low-quality medical images without producing visual artefacts. Also, the performance of the proposed method is consistent for multiple modalities. In every pair, left: input image and right: Enhanced Image. }
\label{realData}
\end{figure*}

\subsection{Real-world Medical Image Enhancement}
Apart from performing enhancement on the synthesised image samples, the feasibility of the proposed method has also studied with real-world perceptually degraded images. Here, real-world images have been extracted from the collected data samples. Fig. \ref{realData} illustrates the performance of the proposed method in enhancing real-world medical images. It can be seen that the proposed method can drastically improve the details of the low-quality real-world medical image and produce visually plausible images. The perceptual enhancement obtained by the proposed method is natural-looking and does not possess any disturbing artefacts. Also, the performance of the proposed method is data-independent and reveals the practicability of the proposed method in multidisciplinary medical image enhancement. 

\subsection{Medical Image Denoising}

\begin{figure*}[!htb]
\centering
\includegraphics[width=15cm,keepaspectratio]{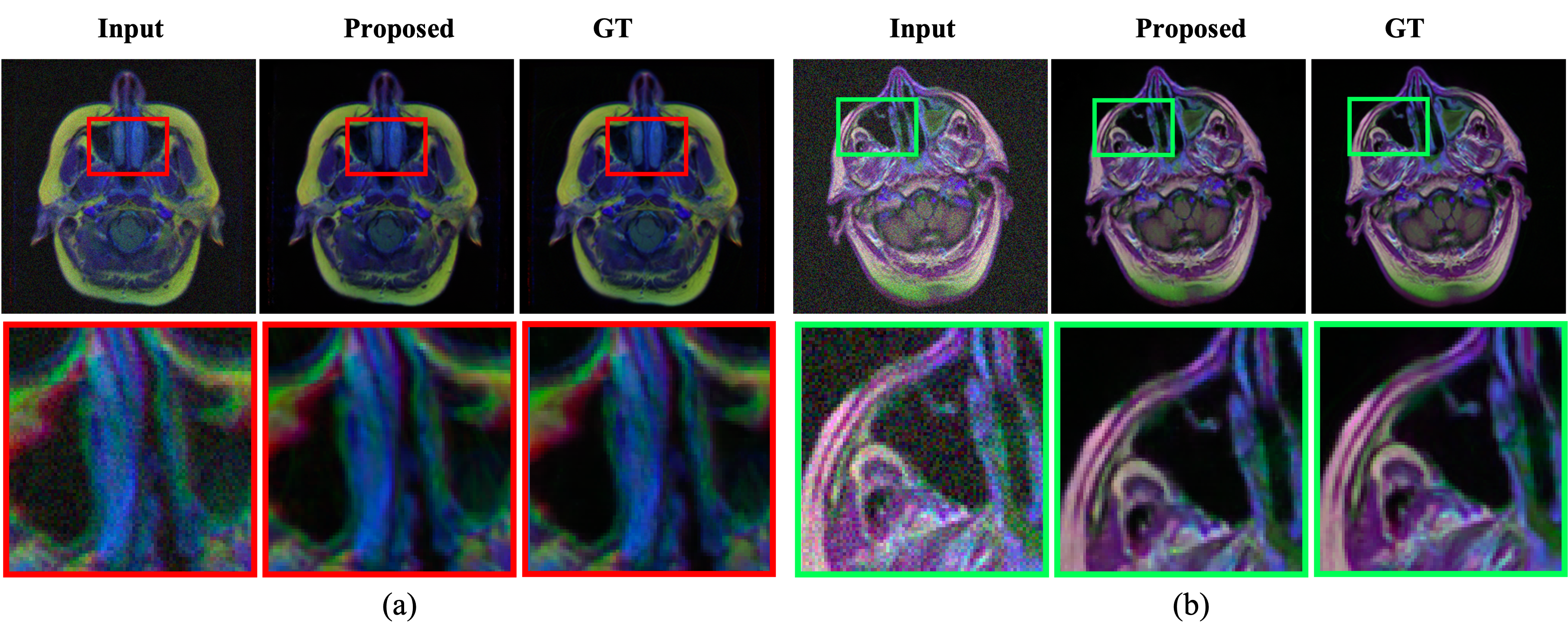}
\caption{ Denoising performance of proposed method. It can suppress image noise without producing any visual artefacts. (a) Denoising at $\sigma = 25$. (b) Denoising at $\sigma = 50$ }
\label{denoising}
\end{figure*}

Image noise is considered one of the most common drawbacks of medical images. This study comprehensively addresses image denoising with perceptual enhancement. However, considering the widespread impact of medical image denoising, the feasibility of the proposed method has been separately studied. Therefore, the proposed network has been retrained with noisy medical images contaminated with noise sigma[0~50]. Fig. \ref{denoising} illustrates the denoising performance of the proposed method on medical images. It can be observed that the proposed network can suppress medical image noises without producing any visual artefacts. It illustrates the applicability of the proposed deep network for medical image denoising also.

\subsection{Ablation Study}

The consequence of the proposed deep components such as residual block, convolutional gate, and perceptual optimisation have been studied by eliminated from the network architecture. Later, each module and loss have been injected into the network architecture to verify their contributions to the final results. The performance of the network variants has been calculated over the images samples used in the section and summarized with evaluation matrics. Table. \ref{tabAB} illustrates the performance of different network variants for medical image enhancement.


\begin{table*}[!htb]
\caption{Ablation study with different network variants. The proposed components accelerate the learning process for medical image enhancement.}
\centering
\scalebox{.7}{\begin{tabular}{lllllllllllll}
\toprule[0.4pt]\toprule[0.4pt]
\multirow{2}{*}{\textbf{Net. Variant}} & \multicolumn{4}{l}{\textbf{Module}}                                                 & \multicolumn{2}{l}{\textbf{Radiology Images}} & \multicolumn{2}{l}{\textbf{Dermatology Image}} & \multicolumn{2}{l}{\textbf{Microscopic Images}} & \multicolumn{2}{l}{\textbf{Average}} \\
                                       & \textbf{Res. Block} & \textbf{Feat. Gate} & \textbf{Feat. Loss} & \textbf{Adversarial loss} & \textbf{PSNR}         & \textbf{DeltaE}       & \textbf{PSNR}         & \textbf{DeltaE}        & \textbf{PSNR}          & \textbf{DeltaE}        & \textbf{PSNR}    & \textbf{DeltaE}   \\ \toprule
Base                                   &   \xmark                  & \xmark                        &      \xmark               &     \xmark              & 25.33                 & 5.71                  & 18.97                 & 10.16                  & 27.05                  & 1.77                   & 23.78            & 5.88              \\
Base + Res. block                                    &     \cmark                &      \xmark                 &       \xmark                &      \xmark               & 26.93                 & 5.12                  & 19.61                 & 8.74                   & 29.02                  & 1.49                   & 25.19            & 5.17              \\
Base + Res. block + Gate                                   &    \cmark                 &     \cmark                &        \xmark             &        \xmark           & 27.66                 & 3.74                  & 20.43                 & 8.56                   & 29.79                  & 1.34                   & 25.96            & 4.55              \\
Base + Res. block + Gate + Reg. Feat. Loss                                &        \cmark               &   \cmark                    &  \cmark                     &       \xmark            & 28.29                 & 3.43                  & 21.44                 & 7.5                    & 29.78                  & 1.25                   & 26.50            & 4.06              \\
\textbf{Proposed}                      &     \cmark                  &           \cmark            &       \cmark                &    \cmark                 & \textbf{29.04}        & \textbf{3.21}         & \textbf{23.11}        & \textbf{6.36}          & \textbf{30.69}         & \textbf{1.11}          & \textbf{27.61}   & \textbf{3.56}  \\ \toprule[0.4pt]\toprule[0.4pt]  
\end{tabular}}
\label{tabAB}
\end{table*}

\begin{figure*}[!htb]
\centering
\includegraphics[width=15cm,keepaspectratio]{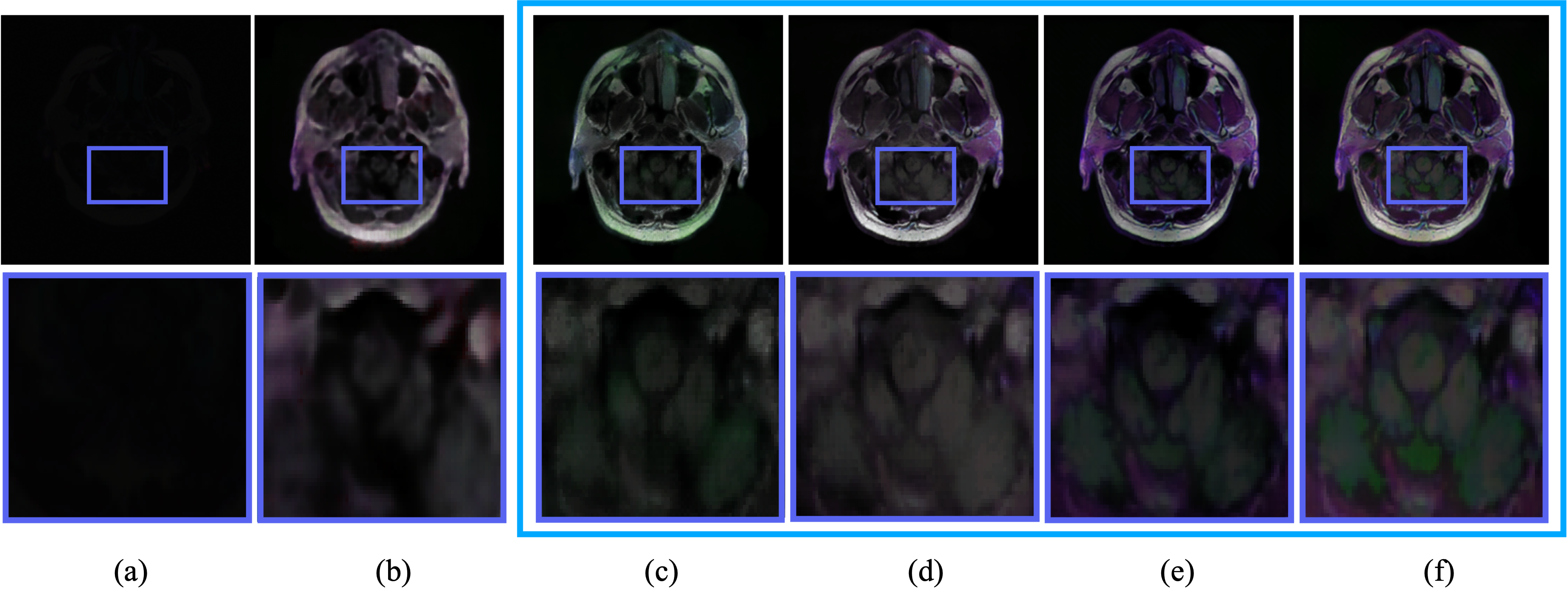}
\caption{ Visual comparison with network variants. Residual block helps the model in recovering details, gating mechanism helps in refining learned features, feature loss enhances visual appearance, and GAN helps in recovering color information in complex structures. (a) Input (low-quality) image. (b) Base network, (c) Base + residual block. (d) Base + residual block + feature gate. (e) Base + residual block + feature gate + regularized feature loss. (f) Proposed network. }
\label{visAB}
\end{figure*}

Apart from quantitative evaluation, the performance of network variants has been confirmed with a visual comparison. Fig. \ref{visAB} depicts the contribution of the proposed components to enhancing low-quality medical images. It can be seen that the residual layers help the proposed deep network to recover details, and the contextual getting mechanism helps the network to refine the learned information to maximize the learning process. Additionally, regularized feature loss can allow the network to produce perceptually plausible images, whereas adversarial loss helps the network to recover colour information in complex respiratory structures.

\section{Discussion}

\begin{figure}[!htb]
\centering
\includegraphics[width=9cm,keepaspectratio]{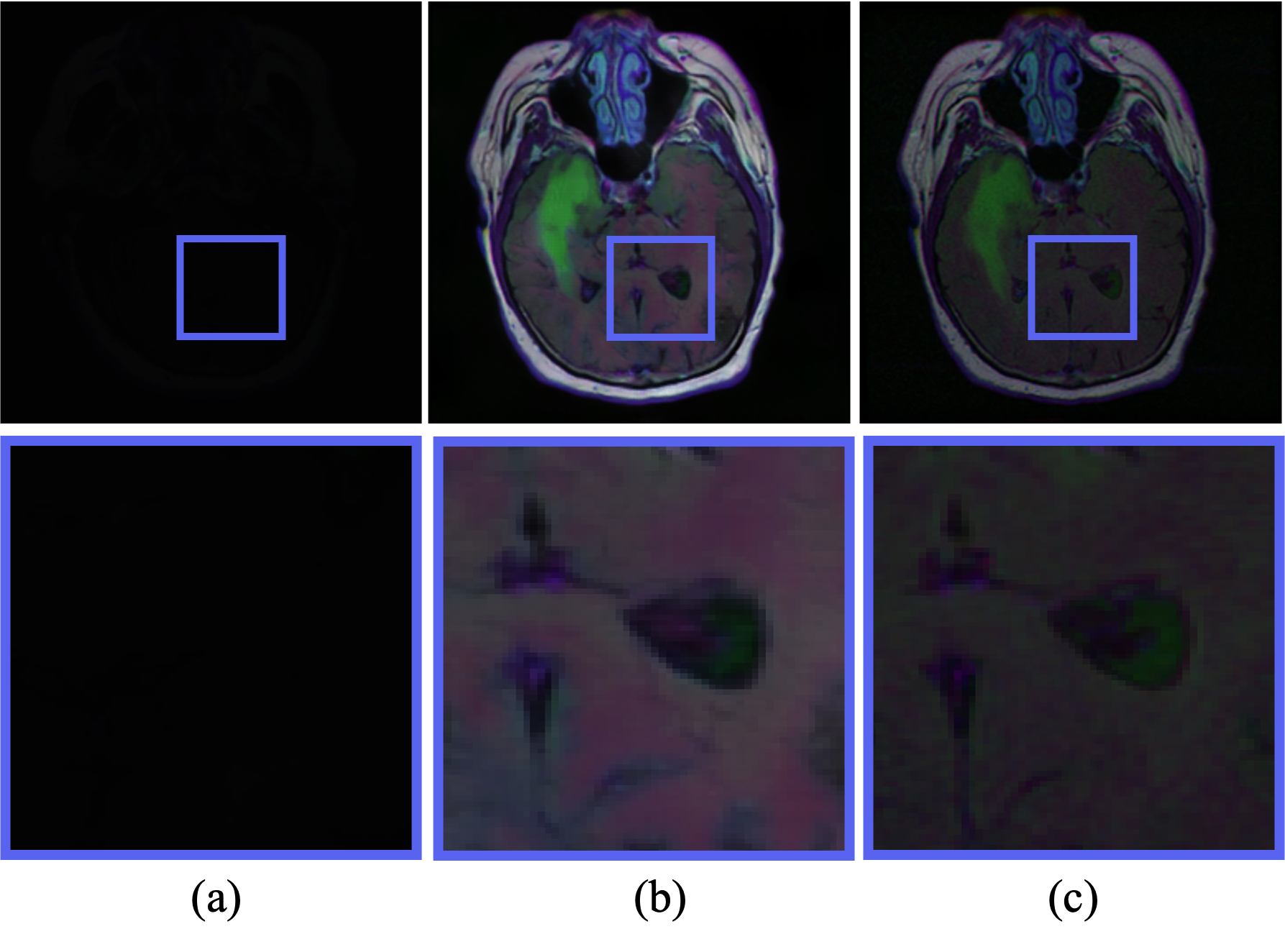}
\caption{ Limitation of proposed method. In some cases, the model can produce brighter images compare to its groundtruth image. }
\label{failure}
\end{figure}

The proposed method collected medical images of different modalities and simulated the low-quality images for pair-wise training. Despite being trained with synthesized pair samples, the proposed model can handle real-world low-quality medical images without showing any artefacts. However, the performance of the proposed method can vary in extreme real-world circumstances (i.e., low-quality images with motion blurs). In some cases, the proposed model can  also produce brighter images, as shown in fig. \ref{failure}.

The proposed method has generalized its performance by combining heterogeneous images collected from different open medical image sources. The proposed model can be trained and optimized by leveraging federated learning \cite{yang2019federated, bonawitz2019towards}. Thus, the model can leverage multiple decentralized edge devices to train with diverse data samples and address the data dependency. The proposed method also could be extended with an unpair training strategy.

A complete study for unveiling the feasibility of the proposed enhancement method in 3D imaging space could be a promising future direction. Therefore, the VGG network utilized for feature loss and the proposed network architecture needs to be restructured for handling 3D inputs.  It is worth noting that the VGG network is a deep convolutional network that can be divided into three parts: input layer, feature extractor, and fully connected layers. Here, the feature extraction portion is fully convolutional and only used for the loss calculation. Typically, the VGG based loss functions modified the input and output layers as per the requirement of the target application \cite{ran2019denoising}. It has been planned to restructure the VGG loss in the foreseeable future to study the applicability of the proposed method for 3D medical images. Also, the feasibility of a deep perceptual enhancement method for video analysis is still due.

\section{Conclusion}
In this study, a novel learning-based deep method has been proposed to enhance the low-quality medical images for enhancing CAD applications and diagnosis process.  The proposed study incorporates a deep model comprising a residual block and gatting mechanism in an encoder-decoder structure to reduce the visual artefacts.  Also, this study has introduced a multi-term objective function to produce natural-looking enhanced images. The feasibility of the proposed study has extensively studied with distinctive experiments. The experiment results illustrated that the proposed method outperforms the existing image enhancement method in qualitative and quantitative evaluation.  The experimental results also demonstrate that the proposed method can drastically improve the performance of the medical image analysis task and can substantially improve the real-world low-quality medical images. Overall, the proposed method reveals a new dimension and new possibilities in medical imaging, explicitly in medical image analysis. It has been planned to extend the proposed study with 3D medical images in the foreseeable future.

\section*{Acknowledgment}
This work was supported in part by the National Research Foundation of Korea (NRF) grant funded by the Korean government (MSIT) (NRF-2021R1A2C1014432) and in part by the NRF grant funded by the Ministry of Science and ICT (MSIT) through the Development Research Program (NRF-2022R1G1A101022611).



\end{document}